\documentclass[12pt,prd,showpacs,tightenlines,nofootinbib]{revtex4}
\usepackage{bm}
\usepackage{graphics}
\usepackage{rotating}
\usepackage{epsfig}
\begin{document}
\begin{flushright}{HU-EP-09/53}\end{flushright}
\title{
Heavy-light meson spectroscopy and Regge trajectories
in the relativistic quark model}
\author{D. Ebert$^{1}$, R. N. Faustov$^{1,2}$  and V. O. Galkin$^{1,2}$}
\affiliation{
$^1$ Institut f\"ur Physik, Humboldt--Universit\"at zu Berlin,
Newtonstr. 15, D-12489  Berlin, Germany\\
$^2$ Dorodnicyn Computing Centre, Russian Academy of Sciences,
  Vavilov Str. 40, 119991 Moscow, Russia}

\begin{abstract}
Masses of the ground, orbitally and radially excited  states of heavy-light
mesons are calculated within the framework of the QCD-motivated relativistic quark
model based on the quasipotential approach. Both light ($q=u,d,s$) and
heavy ($Q=c,b$)
quarks are treated fully relativistically without application of the
heavy quark $1/m_Q$ expansion. The Regge trajectories  in the ($M^2$,$J$)
and ($M^2$, $n_r$) planes are investigated and their parameters are
obtained. The results are in good agreement with available
experimental data except for the masses of the anomalous
$D^*_{s0}(2317)$, $D_{s1}(2460)$ and $D_{sJ}^*(2860)$ states. 
\end{abstract}


\maketitle

\section{Introduction}
\label{sec:intr}

Recently significant experimental progress has been achieved in
studying the spectroscopy of mesons with one heavy ($Q=c,b$) and one
light ($q=u,d,s$) quarks \cite{pdg}. Several new excited states of
heavy-light mesons were 
discovered, some of which have rather unexpected properties \cite{dsexp,dexp}. 

The most
investigated and intriguing issue is the charmed-strange meson sector, where
masses of nine mesons have been measured
\cite{pdg,dsexp,ds2710,ds3040}. Even six years after the 
discovery of $D^*_{s0}(2317)$ and $D_{s1}(2460)$ mesons their
nature remains controversial in the literature. The abnormally light
masses of these mesons put them below $DK$ and $D^*K$ thresholds thus
making these states narrow since the only allowed  decays
violate isospin. The peculiar feature of these mesons is that they
have masses almost equal or even lower than the masses of their
charmed counterparts  $D^*_0(2400)$ and $D_1(2427)$ \cite{dexp,dsexp,pdg}. Most of the
theoretical approaches including lattice QCD \cite{lattice}, QCD sum
rule \cite{sr} and
different quark model \cite{hlmass,qm} calculations give masses of the $0^+$
and $1^+$ $P$-wave $c\bar s$ states significantly heavier (by 100-200 MeV) than
the measured ones. Different theoretical solutions of this problem
were proposed \cite{sw} including consideration of these mesons as 
chiral partners of $0^-$ and $1^-$ states \cite{chiral}, $c\bar s$ states which are
strongly influenced by the nearby $DK$ thresholds \cite{cs}, $DK$ or
$D_s\pi$ molecules \cite{mol},
a mixture of $c\bar s$ and tetraquark states \cite{tetr}. However the universal
understanding of their nature is still missing. Therefore it is
very important to observe their bottom counterparts. The unquenched lattice
calculations of their masses can be found in Ref.~\cite{tmc}

Very recently three new charmed-strange mesons $D_{s1}(2710)$,
$D_{sJ}^*(2860)$  and $D_{sJ}(3040)$  were
observed \cite{ds2710,ds3040}. These states are considered to be
candidates for the $2S$, $1D$ and $2P$ states, respectively. Therefore
it is important to have theoretical predictions not only for the
lowest orbital and radial excitations of heavy-light mesons but also
for the highly excited states.

In Refs.~\cite{hlmass} we calculated the masses of ground and first
orbitally and radially excited states of heavy-light mesons  on the
basis of a three-dimensional relativistic wave equation with a
QCD-motivated potential. The heavy quark $1/m_Q$ expansion was used to
simplify calculations, while the dynamics of light quark was treated
fully relativistically. It was found that the heavy quark $1P$
multiplets with total angular momenta of light quark $j=1/2$ ($0^+$,
$1^+$) and $j=3/2$ ($1^+$, $2^+$) are inverted in the infinitely heavy
quark limit.  The account of the first order $1/m_Q$ corrections
results in splittings and shifts of the levels in these multiplets,
which begin to overlap. As a result a very complicated pattern of
$P$-level structure emerges. 

During the last few years we further developed our model for the treatment of
mesons composed from light quarks~\cite{lmes,lregge}. For this
purpose an approach which 
allows to consider  the highly relativistic dynamics of light quarks
without either the $v/c$ or $1/m_q$ expansion was developed.
The consistent 
relativistic treatment of the light quark dynamics resulted in a
nonlinear dependence of the bound state equation on the meson mass
which allowed to get  correct values of pion and kaon masses in
the model \cite{lmes} with explicitly broken chiral symmetry. The obtained
wave functions of 
the pion and kaon were successfully applied to the relativistic
calculation of their decay constants and electromagnetic form factors
\cite{lmes}. Such approach allowed us to get masses of highly
excited light mesons and on this basis to check the linearity and
parallelism of arising Regge trajectories \cite{lregge}. Good overall
agreement of the obtained predictions and experimental data was
found. 

Here we improve and extend our study of heavy-light meson spectroscopy
by using the fully relativistic approach without the heavy quark $1/m_Q$
expansion. We calculate the masses of highly orbitally and radially
excited states and investigate the Regge trajectories both in the
($M^2$,$J$) and ($M^2$, $n_r$) planes ($M$ is the mass, $J$ is the spin and
 $n_r$ is the radial quantum number of the meson state). Such analysis
 is important for 
 elucidating the nature of current and future experimentally observed
 heavy-light mesons.

\section{Relativistic quark model}
\label{sec:rqm}

  In the relativistic quark model based on the quasipotential approach
  a meson is described by the wave 
function of the bound quark-antiquark state, which satisfies the
quasipotential equation  of the Schr\"odinger type~\cite{efg}
\begin{equation}
\label{quas}
{\left(\frac{b^2(M)}{2\mu_{R}}-\frac{{\bf
p}^2}{2\mu_{R}}\right)\Psi_{M}({\bf p})} =\int\frac{d^3 q}{(2\pi)^3}
 V({\bf p,q};M)\Psi_{M}({\bf q}),
\end{equation}
where the relativistic reduced mass is
\begin{equation}
\mu_{R}=\frac{E_1E_2}{E_1+E_2}=\frac{M^4-(m^2_1-m^2_2)^2}{4M^3},
\end{equation}
and $E_1$, $E_2$ are given by
\begin{equation}
\label{ee}
E_1=\frac{M^2-m_2^2+m_1^2}{2M}, \quad E_2=\frac{M^2-m_1^2+m_2^2}{2M}.
\end{equation}
Here $M=E_1+E_2$ is the meson mass, $m_{1,2}$ are the quark masses,
and ${\bf p}$ is their relative momentum.  
In the center-of-mass system the relative momentum squared on mass shell 
reads
\begin{equation}
{b^2(M) }
=\frac{[M^2-(m_1+m_2)^2][M^2-(m_1-m_2)^2]}{4M^2}.
\end{equation}

The kernel 
$V({\bf p,q};M)$ in Eq.~(\ref{quas}) is the quasipotential operator of
the quark-antiquark interaction. It is constructed with the help of the
off-mass-shell scattering amplitude, projected onto the positive
energy states. 
Constructing the quasipotential of the quark-antiquark interaction, 
we have assumed that the effective
interaction is the sum of the usual one-gluon exchange term with the mixture
of long-range vector and scalar linear confining potentials, where
the vector confining potential
contains the Pauli interaction. The quasipotential is then defined by
  \begin{equation}
\label{qpot}
V({\bf p,q};M)=\bar{u}_1(p)\bar{u}_2(-p){\mathcal V}({\bf p}, {\bf
q};M)u_1(q)u_2(-q),
\end{equation}
with
$${\mathcal V}({\bf p},{\bf q};M)=\frac{4}{3}\alpha_sD_{ \mu\nu}({\bf
k})\gamma_1^{\mu}\gamma_2^{\nu}
+V^V_{\rm conf}({\bf k})\Gamma_1^{\mu}
\Gamma_{2;\mu}+V^S_{\rm conf}({\bf k}),$$
where $\alpha_s$ is the QCD coupling constant, $D_{\mu\nu}$ is the
gluon propagator in the Coulomb gauge,
and ${\bf k=p-q}$; $\gamma_{\mu}$ and $u(p)$ are 
the Dirac matrices and spinors.

The effective long-range vector vertex is
given by
\begin{equation}
\label{kappa}
\Gamma_{\mu}({\bf k})=\gamma_{\mu}+
\frac{i\kappa}{2m}\sigma_{\mu\nu}k^{\nu},
\end{equation}
where $\kappa$ is the Pauli interaction constant characterizing the
anomalous chromomagnetic moment of quarks. Vector and
scalar confining potentials in the nonrelativistic limit reduce to
\begin{eqnarray}
\label{vlin}
V^V_{\rm conf}(r)&=&(1-\varepsilon)(Ar+B),\nonumber\\ 
V^S_{\rm conf}(r)& =&\varepsilon (Ar+B),
\end{eqnarray}
reproducing 
\begin{equation}
\label{nr}
V_{\rm conf}(r)=V^S_{\rm conf}(r)+V^V_{\rm conf}(r)=Ar+B,
\end{equation}
where $\varepsilon$ is the mixing coefficient. 

All the model parameters have the same values as in our previous
papers \cite{hlmass,efg}.
The constituent quark masses $m_u=m_d=0.33$ GeV, $m_s=0.5$ GeV,
$m_c=1.55$ GeV, $m_b=4.88$ GeV and
the parameters of the linear potential $A=0.18$ GeV$^2$ and $B=-0.3$ GeV
have the usual values of quark models.  The value of the mixing
coefficient of vector and scalar confining potentials $\varepsilon=-1$
has been determined from the consideration of charmonium radiative
decays \cite{efg} and matching heavy quark effective theory (HQET). 
Finally, the universal Pauli interaction constant $\kappa=-1$ has been
fixed from the analysis of the fine splitting of heavy quarkonia ${
}^3P_J$- states \cite{efg}. In this case, the long-range chromomagnetic
interaction of quarks, which is proportional to $(1+\kappa)$, vanishes
in accordance with the flux-tube model.

The quasipotential (\ref{qpot})  can in principal  be used for arbitrary quark
masses.  The substitution 
of the Dirac spinors  into (\ref{qpot}) results in an extremely
nonlocal potential in the configuration space. Clearly, it is very hard to 
deal with such potentials without any additional approximations.
 In order to simplify the relativistic $q\bar q$ potential, we make the
following replacement in the Dirac spinors:
\begin{equation}
  \label{eq:sub}
  \epsilon_{1,2}(p)=\sqrt{m_{1,2}^2+{\bf p}^2} \to E_{1,2}
\end{equation}
(see the discussion of this point in \cite{hlmass,lmes}).  This substitution
makes the Fourier transformation of the potential (\ref{qpot}) local.

The resulting $Q\bar q$ potential then reads
\begin{equation}
  \label{eq:v}
  V(r)= V_{\rm SI}(r)+ V_{\rm SD}(r),
\end{equation}
where the explicit expression for the spin-independent $V_{\rm SI}(r)$
can be found in Ref.~\cite{lregge}.
The structure of the spin-dependent potential is given by
\begin{equation}
  \label{eq:vsd}
   V_{\rm SD}(r)=a_1\, {\bf L}{\bf S}_1+a_2\, {\bf L}{\bf S}_2+
b \left[-{\bf S}_1{\bf S}_2+\frac3{r^2}({\bf S}_1{\bf r})({\bf
    S}_2{\bf r})\right]+ c\, {\bf S}_1{\bf S}_2+ d\, ({\bf L}{\bf
  S}_1) ({\bf L}{\bf S}_2),
\end{equation}
where $ {\bf L}$ is the orbital angular momentum, ${\bf S}_i$ is the
quark spin. The coefficients $a_1$, $a_2$, $b$, $c$ and $d$ are expressed
through the corresponding derivatives of the Coulomb and
confining potentials. Their explicit expressions are given in Ref.~\cite{lregge}.  

Since we deal with mesons containing light quarks we adopt for the QCD
coupling constant $\alpha_s(\mu^2)$ the
simplest model with freezing \cite{bvb}, namely
\begin{equation}
  \label{eq:alpha}
  \alpha_s(\mu^2)=\frac{4\pi}{\displaystyle\beta_0
\ln\frac{\mu^2+M_B^2}{\Lambda^2}}, \qquad \beta_0=11-\frac23n_f,
\end{equation}
where the scale is taken as $\mu=2m_1
m_2/(m_1+m_2)$, the background mass is $M_B=2.24\sqrt{A}=0.95$~GeV \cite{bvb}, and
$\Lambda=413$~MeV was fixed from fitting the $\rho$
mass \cite{lregge}. Note that the other popular
parametrization of $\alpha_s$ with freezing \cite{shirkov} leads to close
values.   

The resulting quasipotential equation with the complete kernel
(\ref{eq:v}) is solved numerically without any approximations.

\section{Results and discussion}
\label{sec:rd}

The calculated masses of heavy-light $D$, $D_s$, $B$ and $B_s$ mesons are
given in Tables~\ref{tab:csmm} and~\ref{tab:bsmm} ($n=n_r+1$,
$L$ is the orbital momentum and $S$ is the total spin). They  are confronted
with available experimental data from PDG \cite{pdg}.

{\squeezetable\begin{table}
\caption{Masses of  charmed ($q=u,d$) and charmed-strange mesons
   (in MeV).} 
   \label{tab:csmm}
\begin{ruledtabular}
\begin{tabular}{cccccccc}
\multicolumn{2}{l}{\underline{\phantom{p}\hspace{0.9cm}State\hspace{0.9cm}}}&Theory
&\multicolumn{2}{l}{\underline{\hspace{1.1cm}Experiment\hspace{1.1cm}}}&
Theory&\multicolumn{2}{r}{\underline{\hspace{1.2cm}Experiment\hspace{1.2cm}}}\\
$n^{2S+1}L_J$&$J^{P}$&$c\bar q$ & meson &mass& $c\bar
s$ & meson&mass\\[2pt]
\hline
$1^1S_0$& $0^{-}$&1871& $D$&1869.62(20)&1969& $D_s$ & 1968.49(34)\\
$1^3S_1$& $1^{-}$&2010& $D^*(2010)$&2010.27(17)
&2111&$D^*_s$& 2112.3(5)\\
$1^3P_0$& $0^{+}$&2406&
$D_0^*(2400)$&$\left\{\begin{array}{l}2403(40)(^\pm)\\2352(50)(^0) \end{array}\right.$ &2509&$D_{s0}^*(2317)$&2317.8(6)\\
$1P_1$& $1^{+}$&2469& $D_1(2430)$&2427(40) &2574&$D_{s1}(2460)$&2459.6(6)\\
$1P_1$& $1^{+}$&2426& $D_1(2420)$&2423.4(3.1)&2536&$D_{s1}(2536)$&2535.35(60)\\
$1^3P_2$& $2^{+}$&2460& $D^*_2(2460)$&2460.1($^{+2.6}_{-3.5}$)&2571&$D_{s2}(2573)$&2572.6(9)\\

$2^1S_0$& $0^{-}$&2581& & &2688&&\\
$2^3S_1$& $1^{-}$&2632& $D^*(2637)$& 2637(6)? &2731&$D_{s1}(2710)$& 2710($^{+12}_{-7}$)\\
$1^3D_1$& $1^{-}$&2788& & &2913&&\\
$1D_2$& $2^{-}$&2850& & &2961&&\\
$1D_2$& $2^{-}$&2806& & &2931&&\\
$1^3D_3$& $3^{-}$&2863& & &2971&$D_{sJ}^*(2860)$& 2862($^{+6}_{-3}$)\\
$2^3P_0$& $0^{+}$&2919& & &3054&&\\
$2P_1$& $1^{+}$&3021& & &3154&&\\
$2P_1$& $1^{+}$&2932& & &3067&$D_{sJ}(3040)$ & 3044($^{+30}_{-9}$)\\
$2^3P_2$& $2^{+}$&3012& & &3142&&\\

$3^1S_0$& $0^{-}$&3062& & &3219&&\\
$3^3S_1$& $1^{-}$&3096& & &3242&& \\
$1^3F_2$& $2^{+}$&3090& & &3230&&\\
$1F_3$& $3^{+}$&3145& & &3266&& \\
$1F_3$& $3^{+}$&3129& & &3254&&\\
$1^3F_4$& $4^{+}$&3187& & &3300&&\\

$2^3D_1$& $1^{-}$&3228& & &3383&&\\
$2D_2$& $2^{-}$&3307& &  &3456&&\\
$2D_2$& $2^{-}$&3259& & &3403&&\\
$2^3D_3$& $3^{-}$&3335& & &3469&&\\

$3^3P_0$& $0^{+}$&3346& & &3513&&\\
$3P_1$& $1^{+}$&3461& & &3618&&\\
$3P_1$& $1^{+}$&3365& & &3519&&\\
$3^3P_2$& $2^{+}$&3407& & &3580&&\\

$1^3G_3$& $3^{-}$&3352& & &3508&&\\
$1G_4$& $4^{-}$&3415& & &3554&&\\
$1G_4$& $4^{-}$&3403& & &3546&&\\
$1^3G_5$& $5^{-}$&3473& & &3595&&\\

$4^1S_0$& $0^{-}$&3452& & &3652&&\\
$4^3S_1$& $1^{-}$&3482& & &3669 && \\

$2F_3$& $3^{+}$&3551& & &3710&&\\
$2^3F_4$& $4^{+}$&3610& & &3754&&\\

$2G_4$& $4^{-}$&3791& & &3964&&\\
$2^3G_5$& $5^{-}$&3860& & &4016&&\\

$5^1S_0$& $0^{-}$&3793& & &4033&&\\
$5^3S_1$& $1^{-}$&3822& & &4048&& 
\end{tabular}
\end{ruledtabular}
\end{table}

\begin{table}
\caption{Masses of bottom ($q=u,d$) and bottom-strange mesons
   (in MeV).} 
   \label{tab:bsmm}
\begin{ruledtabular}
\begin{tabular}{cccccccc}
\multicolumn{2}{l}{\underline{\phantom{p}\hspace{0.9cm}State\hspace{0.9cm}}}&Theory
&\multicolumn{2}{l}{\underline{\hspace{1.1cm}Experiment\hspace{1.1cm}}}&
Theory&\multicolumn{2}{r}{\underline{\hspace{1.2cm}Experiment\hspace{1.2cm}}}\\
$n^{2S+1}L_J$&$J^{P}$&$b\bar q$ & meson &mass& $b\bar
s$ & meson&mass\\[2pt]
\hline
$1^1S_0$& $0^{-}$&5280& $B$&5279.5(3)&5372& $B_s$ & 5366.3(6)\\
$1^3S_1$& $1^{-}$&5326& $B^*$&5325.1(5)
&5414&$B^*_s$& 5415.4(1.4)\\
$1^3P_0$& $0^{+}$&5749&$B^*_J(5732)$ & 5698(8)? &5833&&\\
$1P_1$& $1^{+}$&5774& & &5865&$B^*_{sJ}(5850)$&5853(15)?\\
$1P_1$& $1^{+}$&5723& $B_1(5721)$&5723.4(2.0)&5831&$B_{s1}(5830)$&5829.4(7)\\
$1^3P_2$& $2^{+}$&5741& $B^*_2(5747)$&5743(5)&5842&$B_{s2}^*(5840)$&5839.7(6)\\

$2^1S_0$& $0^{-}$&5890& & &5976&&\\
$2^3S_1$& $1^{-}$&5906& & &5992&&\\
$1^3D_1$& $1^{-}$&6119& & &6209&&\\
$1D_2$& $2^{-}$&6121& & &6218&&\\
$1D_2$& $2^{-}$&6103& & &6189&&\\
$1^3D_3$& $3^{-}$&6091& & &6191&&\\
$2^3P_0$& $0^{+}$&6221& & &6318&&\\
$2P_1$& $1^{+}$&6281& & &6345&&\\
$2P_1$& $1^{+}$&6209& & &6321&& \\
$2^3P_2$& $2^{+}$&6260& & &6359&&\\

$3^1S_0$& $0^{-}$&6379& & &6467&&\\
$3^3S_1$& $1^{-}$&6387& & &6475&& \\
$1^3F_2$& $2^{+}$&6412& & &6501&&\\
$1F_3$& $3^{+}$&6420& & &6515&& \\
$1F_3$& $3^{+}$&6391& & &6468&&\\
$1^3F_4$& $4^{+}$&6380& & &6475&&\\

$2^3D_1$& $1^{-}$&6534& & &6629&&\\
$2D_2$& $2^{-}$&6554& &  &6651&&\\
$2D_2$& $2^{-}$&6528& & &6625&&\\
$2^3D_3$& $3^{-}$&6542& & &6637&&\\

$3^3P_0$& $0^{+}$&6629& & &6731&&\\
$3P_1$& $1^{+}$&6685& & &6768&&\\
$3P_1$& $1^{+}$&6650& & &6761&&\\
$3^3P_2$& $2^{+}$&6678& & &6780&&\\

$1^3G_3$& $3^{-}$&6664& & &6753&&\\
$1G_4$& $4^{-}$&6652& & &6762&&\\
$1G_4$& $4^{-}$&6648& & &6715&&\\
$1^3G_5$& $5^{-}$&6634& & &6726&&\\

$4^1S_0$& $0^{-}$&6781& & &6874&&\\
$4^3S_1$& $1^{-}$&6786& & &6879 && \\

$2F_3$& $3^{+}$&6786& & &6880&&\\
$2^3F_4$& $4^{+}$&6784& & &6878&&\\

$2G_4$& $4^{-}$&7007& & &7101&&\\
$2^3G_5$& $5^{-}$&7004& & &7097&&\\

$5^1S_0$& $0^{-}$&7129& & &7231&&\\
$5^3S_1$& $1^{-}$&7133& & &7235&& 
\end{tabular}
\end{ruledtabular}
\end{table}}

The heavy-light meson states  with $J=L$, given in
Tables~\ref{tab:csmm}, \ref{tab:bsmm},
are  mixtures of spin-triplet $|^3L_L\rangle$  and spin-singlet $|^1L_L\rangle$
states:
\begin{eqnarray}
  \label{eq:mix}
  |\Psi_J\rangle&=&|^1L_L\rangle\cos\varphi+|^3L_L\rangle\sin\varphi, \cr
 |\Psi'_J\rangle&=&-|^1L_L\rangle\sin\varphi+|^3L_L\rangle\cos\varphi, \qquad J=L=1,2,3\dots
\end{eqnarray}
where $\varphi$ is a mixing angle and the primed state has the heavier mass.
  Such mixing occurs due to the nondiagonal spin-orbit and
tensor terms in Eq.~(\ref{eq:vsd}). The masses of  physical states were obtained
by diagonalizing the mixing terms. The found values of mixing angle
$\varphi$ are given in Table~\ref{tab:mix}.

In the heavy quark limit heavy-light mesons are usually 
described in the
$|J,j\rangle$ basis, where $j=L+s_q$ is the total angular momentum of the light
quark. The relation between the  $|J,j\rangle$ and $|J,S\rangle$ basises
is given by
\begin{equation}
  \label{eq:js}
   |J;j\rangle=\sum_S(-1)^{J+L+1}\sqrt{(2S+1)(2j+1)}\left\{
   {1/2 \atop L}\ {1/2 \atop J}\ {S\atop j}\right\} |J;S\rangle,
\end{equation}
where $|J;S\rangle$ corresponds to the $|^{2S+1}L_J\rangle$ state.
The following relations for the states with $J=L$ than follow
\begin{eqnarray}
  \label{eq:mixing}
 \left|J=L;j=L+\frac12\right\rangle&=& \sqrt{\frac{L+1}{2L+1}}|J=L;0\rangle
  +\sqrt{\frac{L}{2L+1}}|J=L;1\rangle,\cr
 \left|J=L;j=L-\frac12\right\rangle&=& -\sqrt{\frac{L}{2L+1}}|J=L;0\rangle
  +\sqrt{\frac{L+1}{2L+1}}|J=L;1\rangle.
\end{eqnarray}
In the $m_Q\to\infty$  limit $|\Psi_J\rangle$ and $|\Psi'_J\rangle$ turn into
$\left|J;j=L+\frac12\right\rangle$ and
$\left|J;j=L-\frac12\right\rangle$ states, respectively.
Comparing Eqs.~(\ref{eq:mix}) and (\ref{eq:mixing}) it is easy to obtain 
the infinitely heavy quark limit for to the mixing angle
\begin{equation}
  \label{eq:mqphi}
  \varphi_{m_Q\to\infty}=\arctan\left(\sqrt{\frac{L}{L+1}}\right).
\end{equation}
It is clearly seen from Table~\ref{tab:mix} that the found values of
mixing angles $\varphi$ are very close to $
\varphi_{m_Q\to\infty}$. This means that the physical $|\Psi_J\rangle$ and
$|\Psi'_J\rangle$ states in our model are almost pure 
$\left|J;j=L+\frac12\right\rangle$ and $\left|J;j=L-\frac12\right\rangle$ 
HQET states, respectively. Since $\left|J;j=L+\frac12\right\rangle$
states have higher value of the light quark angular momentum $j$ than
$\left|J;j=L-\frac12\right\rangle$ ones they should decay to the pair
of ground state heavy-light and light mesons in a higher wave and
therefore are expected to be significantly narrower than the partner
states with $j=L-1/2$. For example, the $P$-wave mesons with $j=1/2$
can decay in an $S$-wave and, therefore, are expected to be broad, while
those with  $j=3/2$ can decay  in a $D$-wave and should be
narrow. The found values of the mixing angle $\varphi$ in our model
indicate that there is only a small admixture of broad states to the
narrow ones and therefore they should remain narrow which is in accord
with available experimental data.     
 
\begin{table}
\caption{Mixing angles $\varphi$ for heavy-light mesons (in $^{\circ}$).} 
   \label{tab:mix}
\begin{ruledtabular}
\begin{tabular}{cccccc}
State& $D$ & $D_s$ & $B$ & $B_s$ & $m_Q\to\infty$\\
\hline
$1P$ & 35.5 & 34.5 & 35.0 & 36.0& 35.3\\
$2P$ & 37.5 & 37.6 & 37.3 & 34.0& 35.3\\
$3P$ & 38.4 & 38.2 & 34.1 & 36.7& 35.3\\
$1D$ & 40.7 & 39.2 & 38.0 & 38.1& 39.2\\
$2D$ & 39.0 & 41.2 & 41.6 & 41.1& 39.2\\
$1F$ & 39.6 & 40.5 & 39.5 & 40.1& 40.9\\
$1G$ & 40.2 & 40.3 & 40.4 & 41.9& 41.8\\
\end{tabular}
\end{ruledtabular}
\end{table}

In our analysis we calculated masses of both orbitally and radially excited
heavy-light mesons up to rather high excitation numbers ($L=4$ and
$n_r=4$). This makes it possible 
to construct the heavy-light meson Regge trajectories in the
$(J,M^2)$ and $(n_r,M^2)$ planes. We use the following  definitions. \\ 
a) The $(J,M^2)$ Regge trajectory:

\begin{equation}
  \label{eq:reggej}
J=\alpha M^2+\alpha_0;
\end{equation}

\noindent b) The $(n_r,M^2)$ Regge trajectory:

\begin{equation}
  \label{eq:reggen}
n_r=\beta M^2+\beta_0,
\end{equation}
where $\alpha$, $\beta$ are the slopes and  $\alpha_0$, $\beta_0$ are
intercepts. The relations (\ref{eq:reggej}) and (\ref{eq:reggen})
arise in most models of quark confinement, but with different values
of the slopes.

\begin{figure}
 \includegraphics[width=13cm]{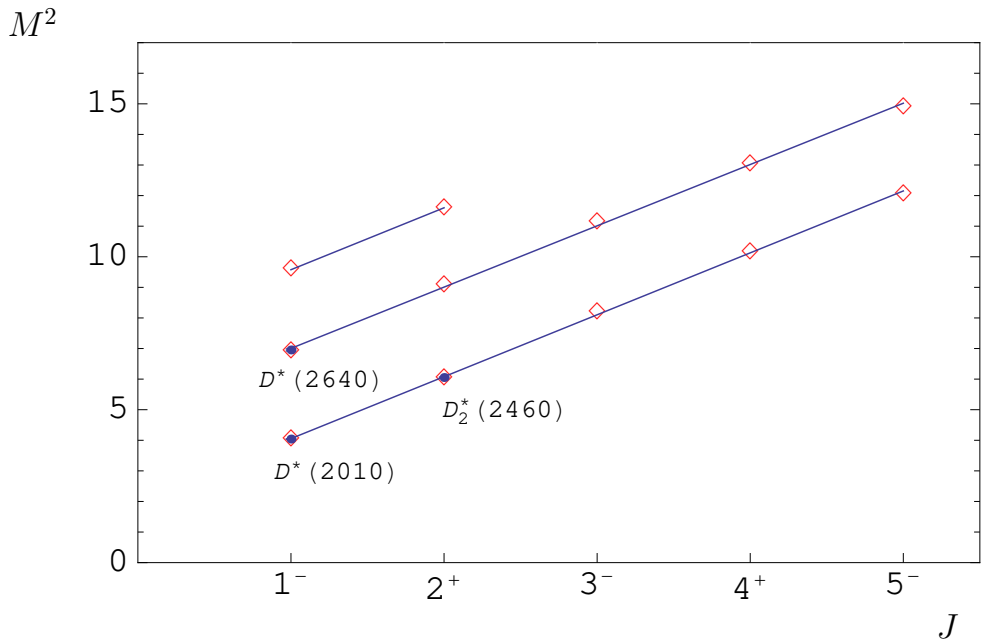} 
\caption{\label{fig:dstar_j} Parent and daughter ($J, M^2$) Regge trajectories for
  charmed mesons  with natural parity. Diamonds are predicted
  masses. Available experimental data are given by dots with  particle
  names; $M^2$ is in GeV$^2$. } 
\end{figure}

\begin{figure}
 \includegraphics[width=13cm]{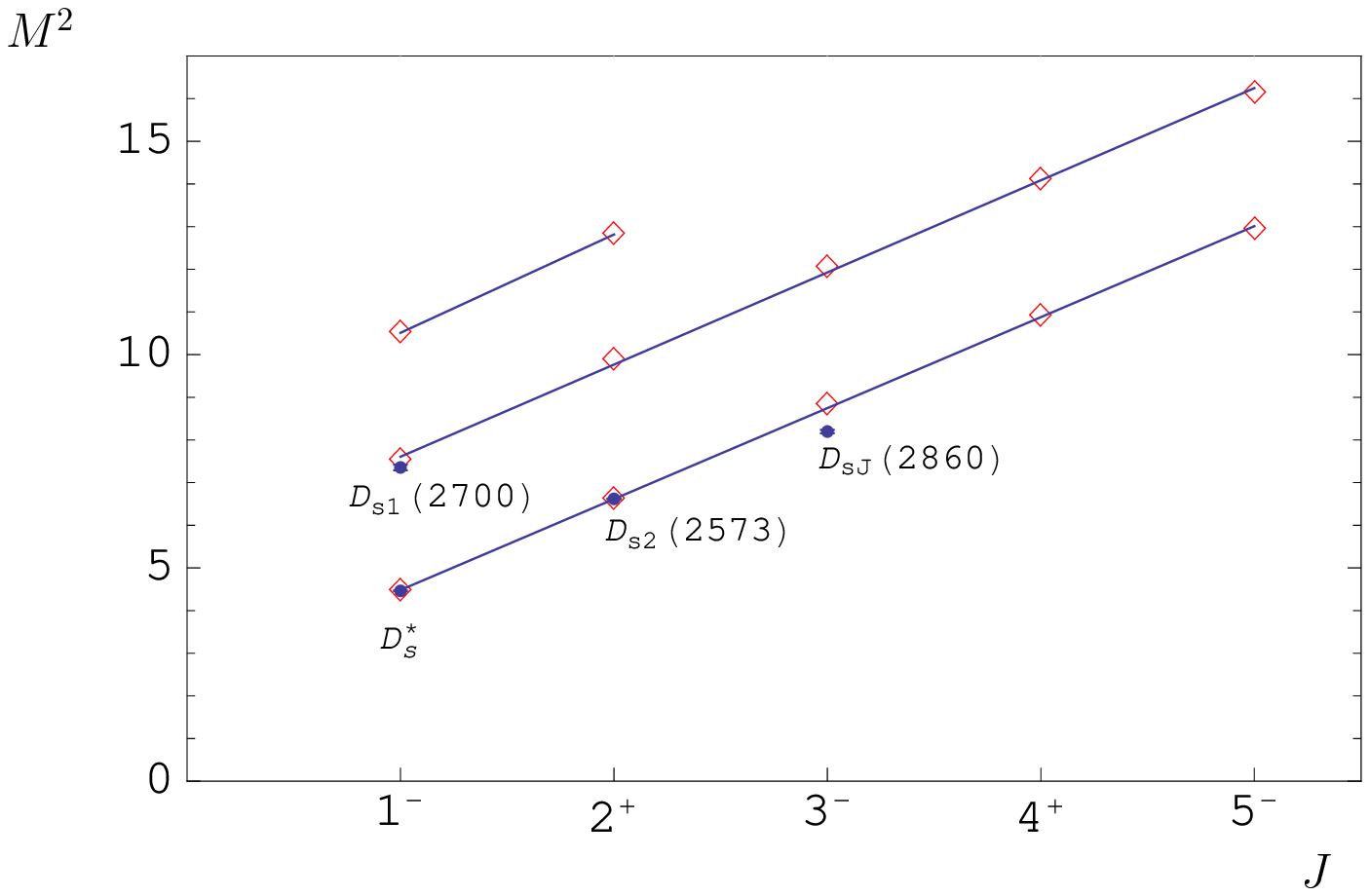}

\caption{\label{fig:dsstar_j} Same as in Fig.~\ref{fig:dstar_j} for
  charmed-strange mesons  with natural parity. }
\end{figure}

\begin{figure}

\includegraphics[width=13cm]{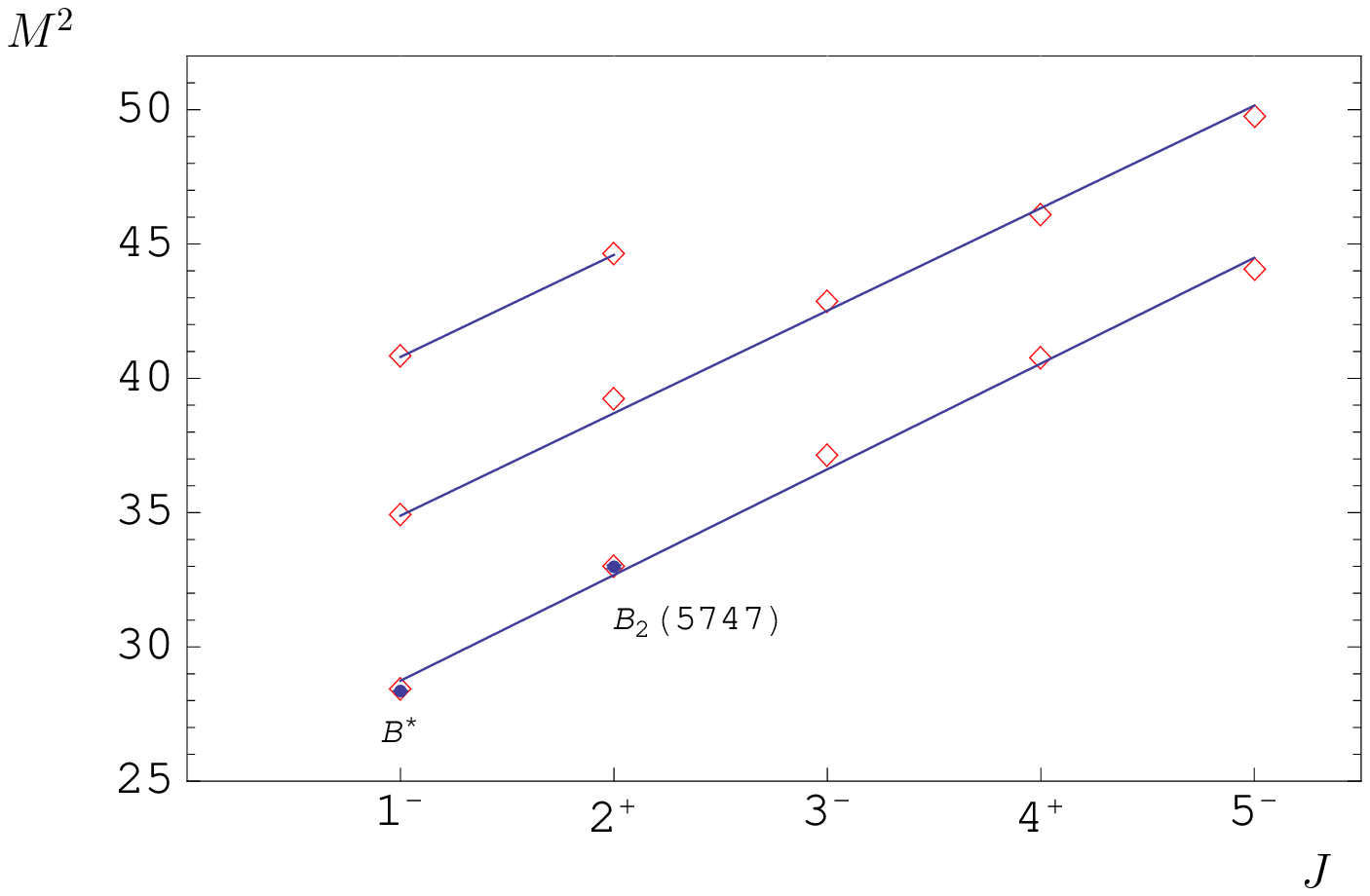} 
\caption{\label{fig:bstar_j} Same as in Fig.~\ref{fig:dstar_j} for
  bottom mesons  with natural parity.  }
\end{figure}

\begin{figure}
 \includegraphics[width=13cm]{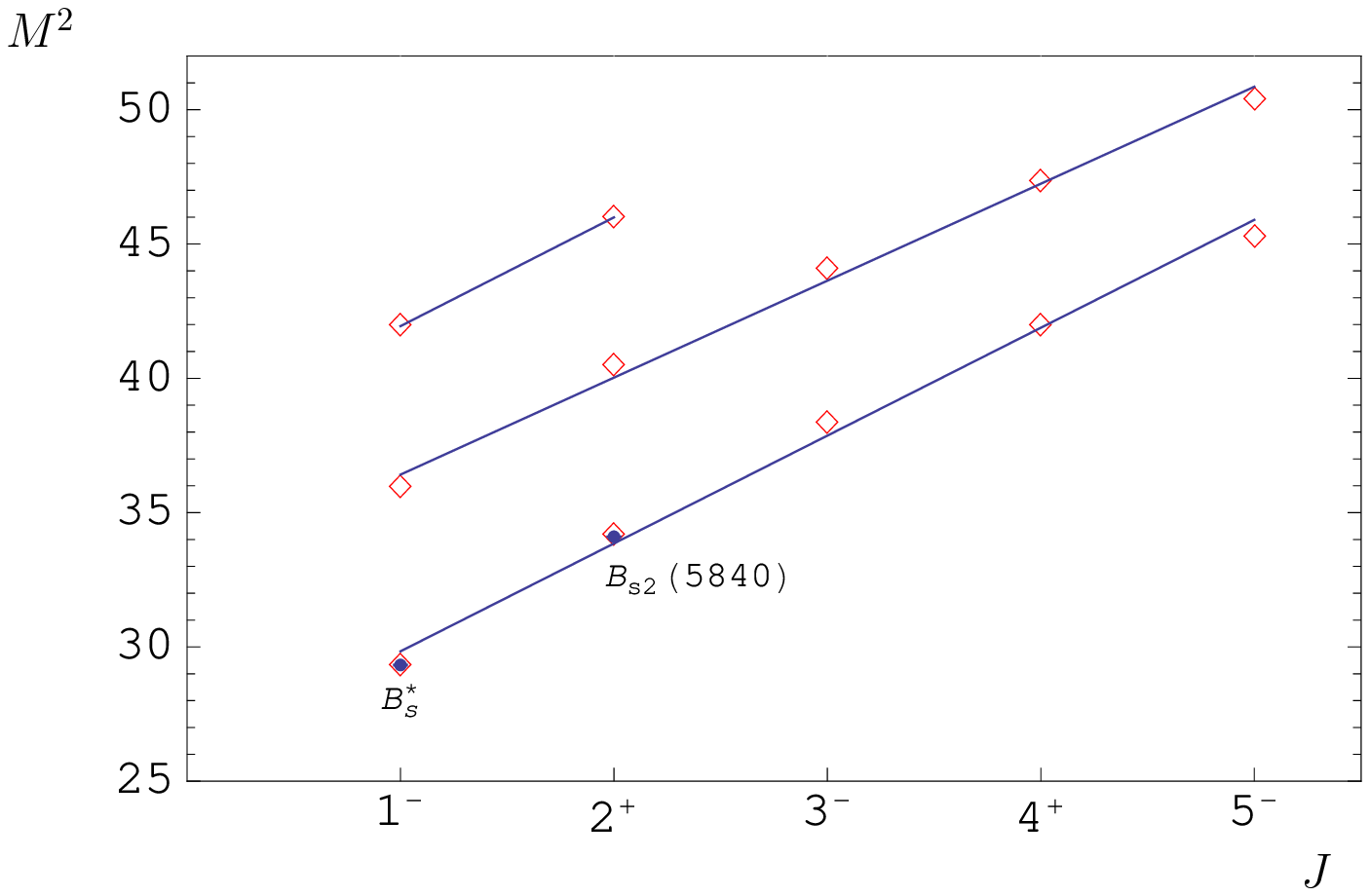}
\caption{\label{fig:bsstar_j} Same as in Fig.~\ref{fig:dstar_j} for
  bottom-strange mesons  with natural parity. }
\end{figure}

\begin{figure}
  \includegraphics[width=13cm]{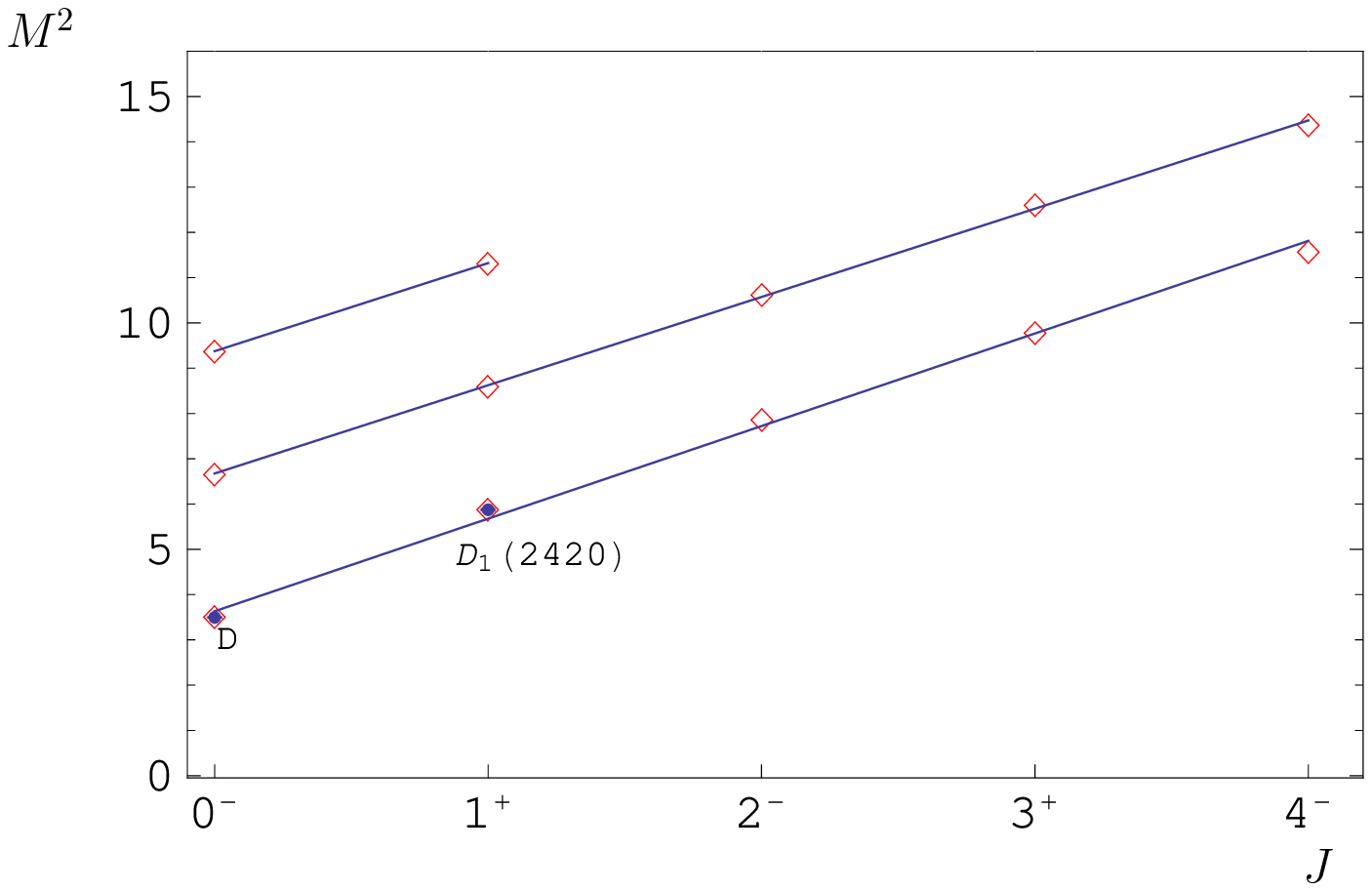}

\caption{\label{fig:d_j} Same as in Fig.~\ref{fig:dstar_j} for
  charmed mesons with unnatural parity. }
\end{figure}

\begin{figure}
  \includegraphics[width=13cm]{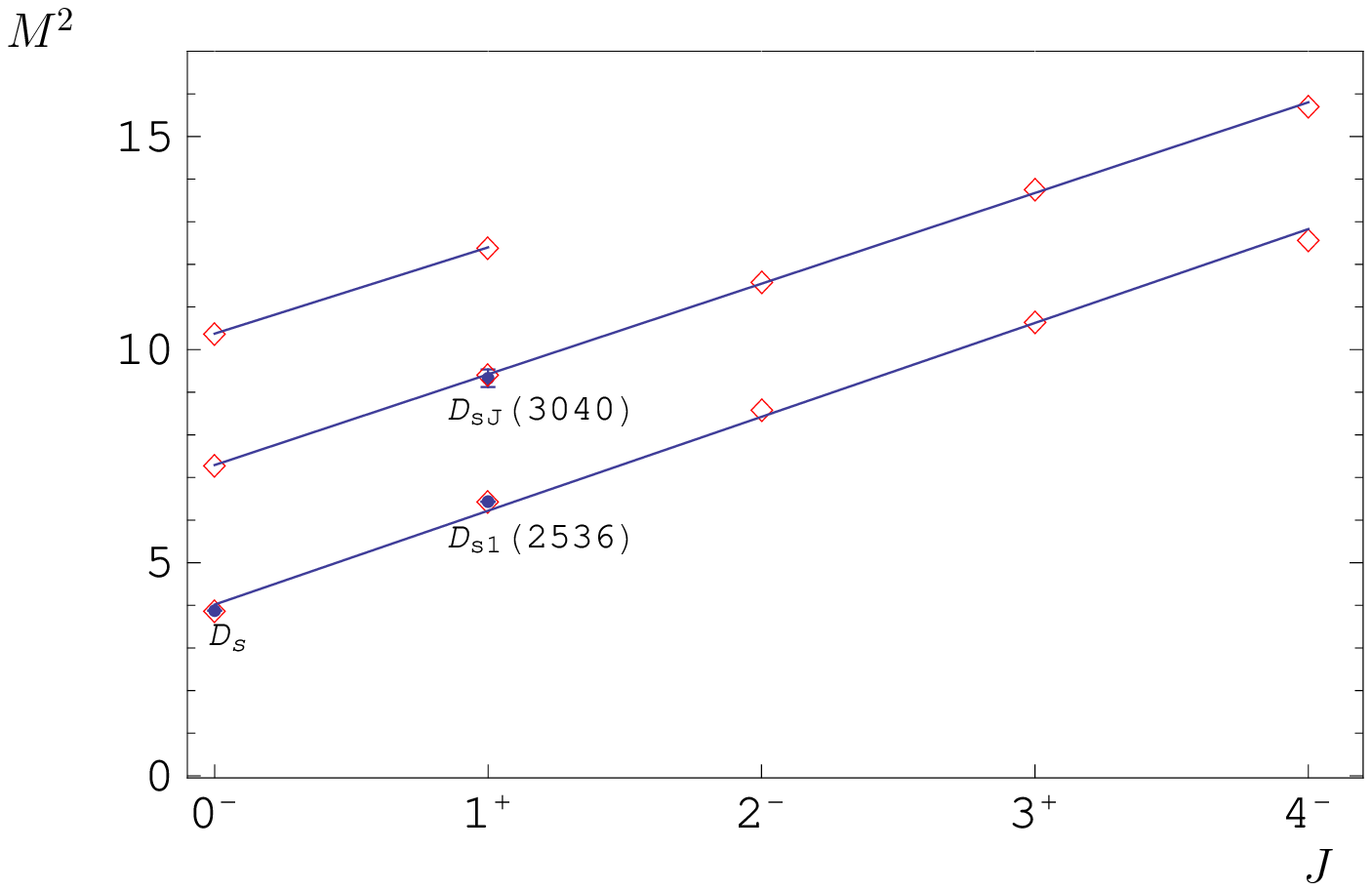}

\caption{\label{fig:ds_j} Same as in Fig.~\ref{fig:dstar_j} for
  charmed-strange mesons with unnatural parity. }
\end{figure}

\begin{figure}
  \includegraphics[width=13cm]{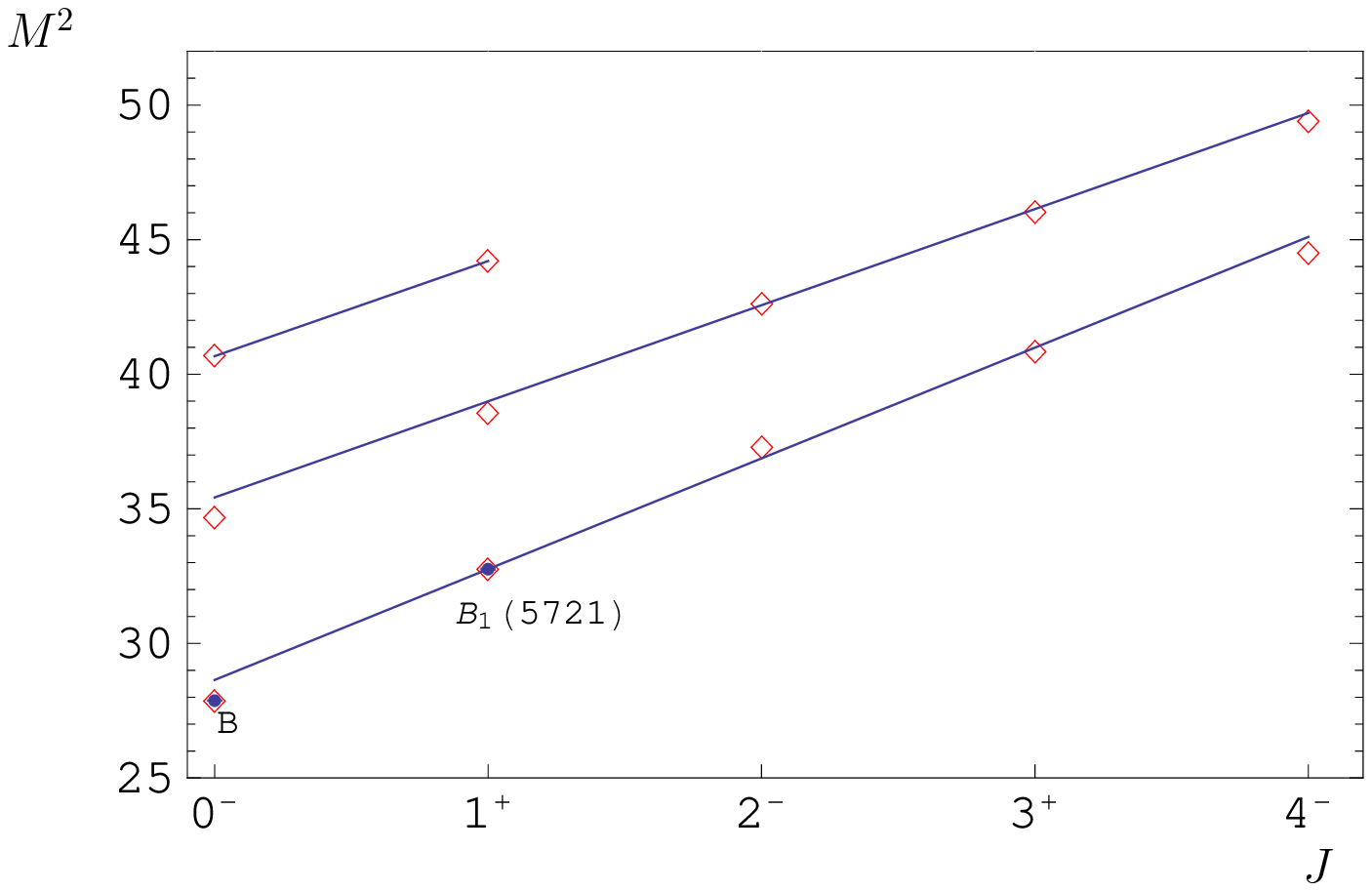}

\caption{\label{fig:b_j} Same as in Fig.~\ref{fig:dstar_j} for
  bottom mesons with unnatural parity. }
\end{figure}

\begin{figure}
  \includegraphics[width=13cm]{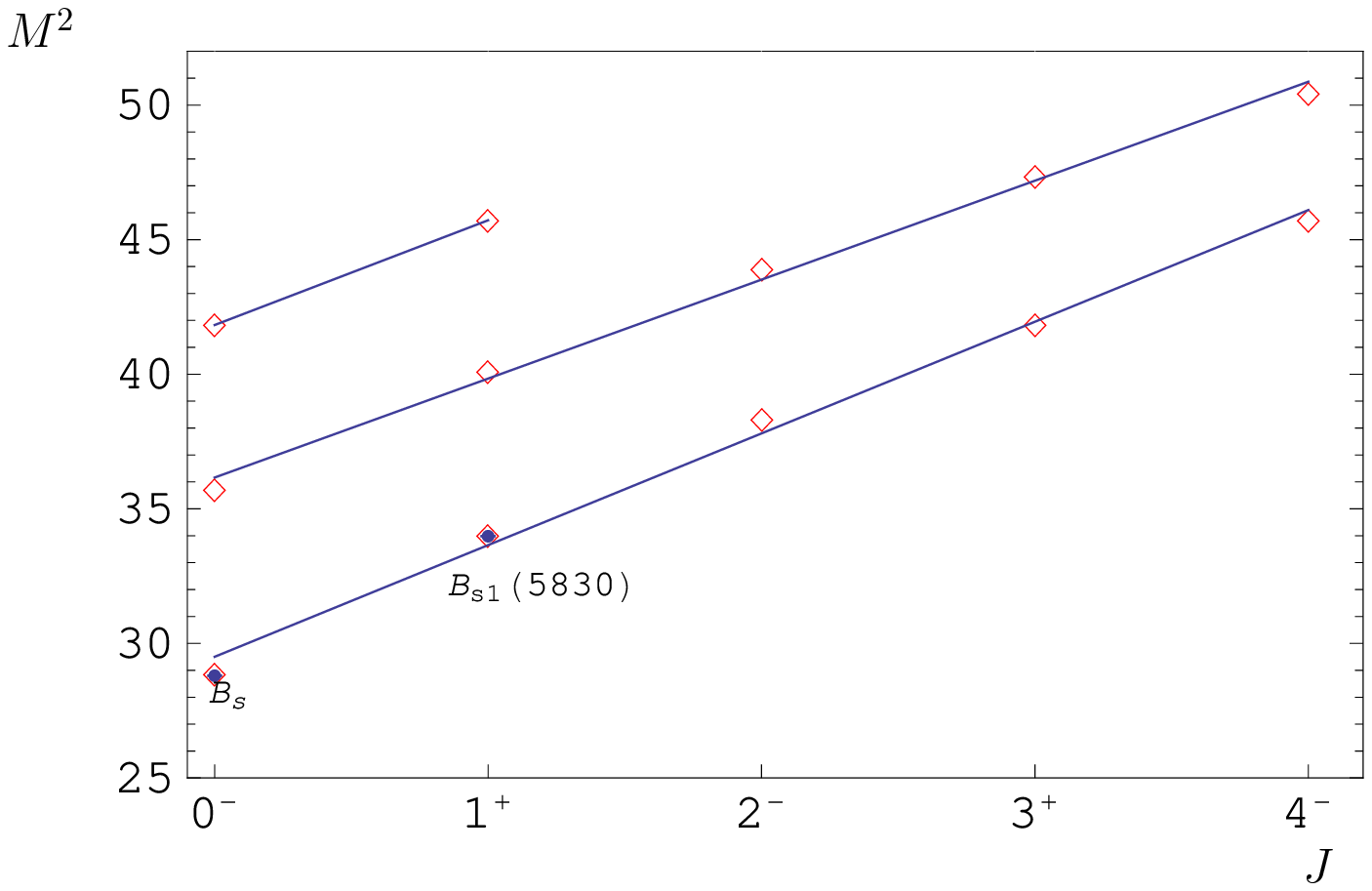}

\caption{\label{fig:bs_j} Same as in Fig.~\ref{fig:dstar_j} for
  bottom-strange mesons with unnatural parity. }
\end{figure}

\begin{figure}
  \includegraphics[width=13cm]{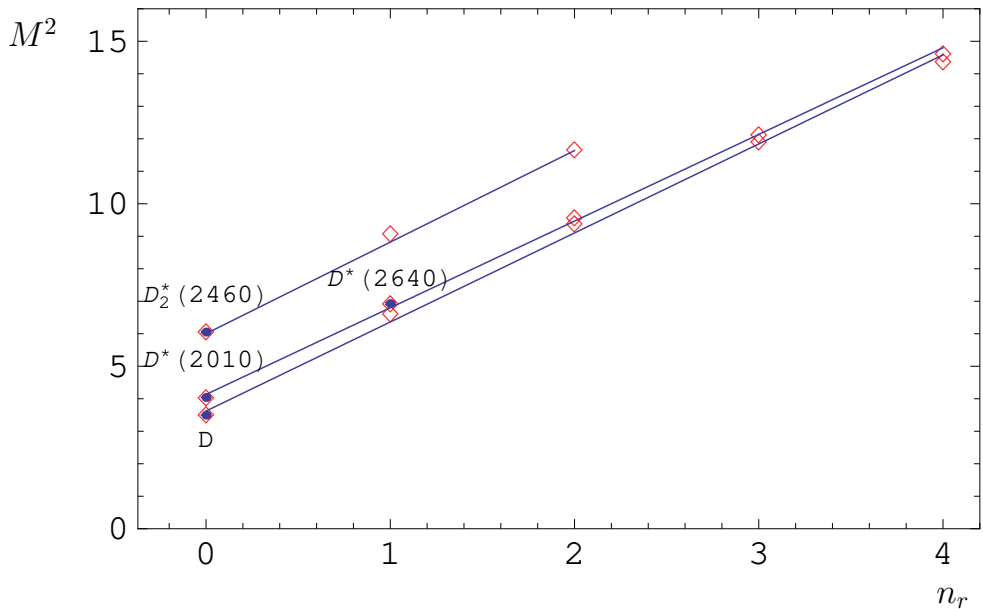}

\caption{\label{fig:d_n} The $(n_r,M^2)$ Regge trajectories for
  pseudoscalar, vector and tensor charmed  mesons (from bottom to
  top). Notations are the same as in Fig.~\ref{fig:dstar_j}. }
\end{figure}

\begin{figure}
  \includegraphics[width=13cm]{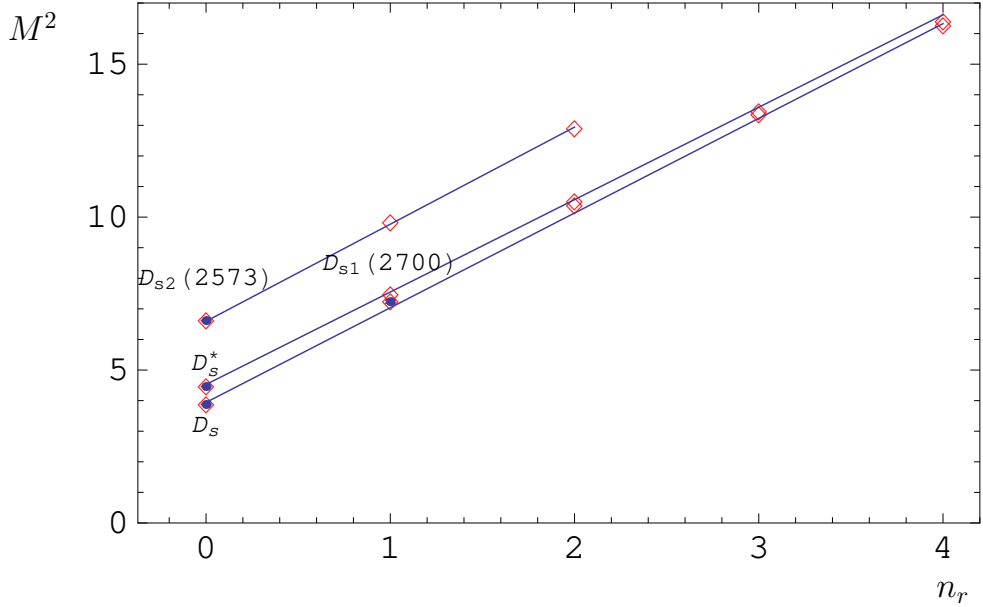}

\caption{\label{fig:ds_n} Same as in Fig.~\ref{fig:d_n} for
  charmed-strange mesons.  }
\end{figure}

\begin{figure}
  \includegraphics[width=13cm]{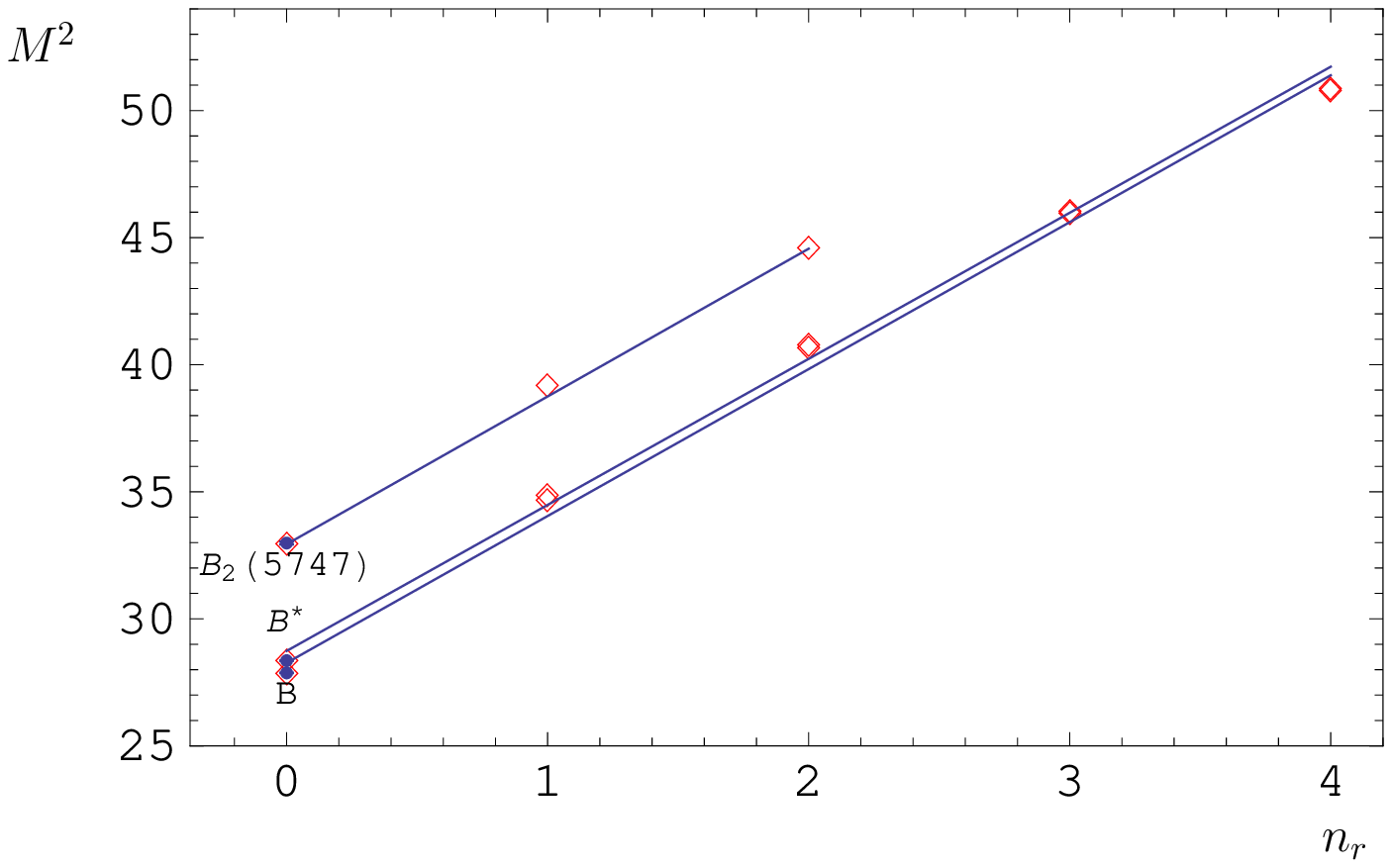}

\caption{\label{fig:bn} Same as in Fig.~\ref{fig:d_n} for
  bottom mesons.  }
\end{figure}

\begin{figure}
  \includegraphics[width=13cm]{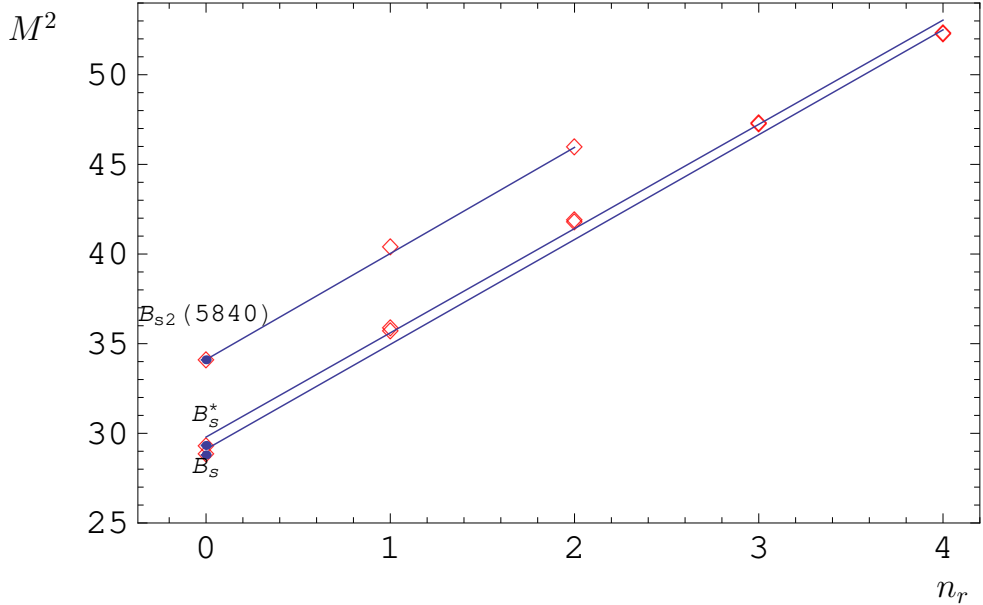}

\caption{\label{fig:bs_n} Same as in Fig.~\ref{fig:d_n} for
  bottom-strange mesons.  }
\end{figure}

In Figs.~\ref{fig:dstar_j}-\ref{fig:bs_j} we plot the Regge trajectories in
the ($J, M^2$) plane for mesons with natural ($P=(-1)^J $) and
unnatural ($P=(-1)^{J-1}$) parity. The Regge trajectories in the
$(n_r,M^2)$ plane are presented 
in Figs.~\ref{fig:d_n}-\ref{fig:bs_n}. The masses calculated in our
model are shown by diamonds. Available experimental data are given by
dots with error bars and corresponding meson names. 
Straight lines were obtained by a
$\chi^2$ fit of the calculated values. The fitted slopes
and intercepts of the Regge trajectories are given in
Tables~\ref{tab:rtj} and \ref{tab:rtn}. We see that the calculated
heavy-light meson masses fit nicely to the linear trajectories in both
planes. These trajectories are almost parallel and equidistant.

\begin{table}
  \caption{Fitted parameters of the $(J,M^2)$ parent and daughter Regge
    trajectories for heavy-light mesons 
    with natural and unnatural parity ($q=u,d$).} 
  \label{tab:rtj}
\begin{ruledtabular}
\begin{tabular}{ccccc}
Trajectory& 
\multicolumn{2}{l}{\underline{\hspace{1.6cm}natural
    parity\hspace{1.6cm}}}&
 \multicolumn{2}{l}{\underline{\hspace{1.6cm}unnatural
    parity\hspace{1.6cm}}}\hspace{-.1cm}\\ 
&$\alpha$ (GeV$^{-2}$)& $\alpha_0$&$\alpha$  (GeV$^{-2}$)&$\alpha_0$\\
\hline
$c\bar q$ &$D^*$&&$D$\\
parent& $0.494\pm0.005$& $-1.003\pm0.040$& $0.489\pm0.016$&
$-1.776\pm0.115$\\
daughter &$0.499\pm0.009$&$-2.495\pm0.091$ & $0.513\pm0.006$&
$-3.424\pm0.063$\\
$c\bar q$ &$D^*_0$&&$D_1$\\
parent& $0.548\pm0.015$& $-3.205\pm0.121$& $0.538\pm0.013$&
$-2.311\pm0.116$\\
daughter &$0.527\pm0.003$&$-4.489\pm0.003$ & $0.557\pm0.005$&
$-4.084\pm0.047$\\
$c\bar s$ &$D^*_s$&&$D_s$\\
parent& $0.469\pm0.004$& $-1.102\pm0.035$& $0.454\pm0.015$&
$-1.824\pm0.114$\\
daughter &$0.463\pm0.008$&$-2.522\pm0.097$ & $0.470\pm0.005$&
$-3.427\pm0.052$\\
$c\bar s$ &$D^*_{s0}$&&$D_{s1}$\\
parent& $0.497\pm0.013$& $-3.161\pm0.119$& $0.482\pm0.007$&
$-2.114\pm0.065$\\
daughter &$0.481\pm0.004$&$-4.489\pm0.047$ & $0.488\pm0.026$&
$-3.793\pm0.326$\\
$b\bar q$ &$B^*$&&$B$\\
parent& $0.254\pm0.010$& $-6.302\pm0.357$& $0.243\pm0.015$&
$-6.960\pm0.572$\\
daughter &$0.282\pm0.012$&$-8.961\pm0.497$ & $0.280\pm0.017$&
$-9.918\pm0.726$\\
$b\bar q$ &$B^*_0$&&$B_1$\\
parent& $0.263\pm0.012$& $-8.774\pm0.474$& $0.262\pm0.013$&
$-7.651\pm0.506$\\
daughter &$0.288\pm0.013$&$-11.232\pm0.590$ & $0.287\pm0.013$&
$-10.135\pm0.575$\\
$b\bar s$ &$B^*_s$&&$B_s$\\
parent& $0.249\pm0.011$& $-6.429\pm0.419$& $0.241\pm0.016$&
$-7.111\pm0.621$\\
daughter &$0.277\pm0.013$&$-9.087\pm0.565$ & $0.272\pm0.012$&
$-9.836\pm0.508$\\
$b\bar s$ &$B^*_{s0}$&&$B_{s1}$\\
parent& $0.259\pm0.013$& $-8.869\pm0.512$& $0.270\pm0.012$&
$-8.258\pm0.495$\\
daughter &$0.285\pm0.013$&$-11.455\pm0.576$ & $0.290\pm0.010$&
$-10.695\pm0.468$\\ 
\end{tabular}
 \end{ruledtabular}
\end{table}

\begin{table}
  \caption{Fitted parameters of the $(n_r,M^2)$  Regge
    trajectories for heavy-light mesons.} 
  \label{tab:rtn}
\begin{ruledtabular}
\begin{tabular}{cccccc}
Meson&$\beta$ (GeV$^{-2}$)& $\beta_0$&Meson&$\beta$ (GeV$^{-2}$)& $\beta_0$\\
\hline
$c\bar q$&&& $c\bar s$\\
$D$& $0.362\pm0.011$& $-1.322\pm0.090$&$D_s$& $0.320\pm0.006$&$-1.273\pm0.053$\\
$D^*$&$0.375\pm0.007$&$-1.550\pm0.058$ &$D_s^*$& $0.334\pm0.002$&$-1.499\pm0.016$\\
$D^*_0$& $0.369\pm0.002$&$-2.141\pm0.018$ &$D^*_{s0}$& $0.331\pm0.001$&$-2.082\pm0.005$\\
$D_1$& $0.339\pm0.006$&$-2.072\pm0.051$&$D_{s1}$&$0.309\pm0.005$& $-2.049\pm0.046$\\
$D_1$& $0.378\pm0.006$&$-2.234\pm0.051$&$D_{s1}$& $0.336\pm0.001$&$-2.161\pm0.001$\\
$D_2$& $0.345\pm0.009$&$-2.101\pm0.082$&$D_{s2}$& $0.313\pm0.004$&$-2.077\pm0.042$\\  
$b\bar q$&&& $b\bar s$\\
$B$& $0.173\pm0.007$& $-4.913\pm0.269$&$B_s$& $0.171\pm0.006$&$-4.978\pm0.249$\\
$B^*$&$0.176\pm0.006$&$-5.082\pm0.243$ &$B_s^*$& $0.172\pm0.006$&$-5.124\pm0.224$\\
$B^*_0$& $0.183\pm0.004$&$-6.069\pm0.151$ &$B^*_{s0}$& $0.177\pm0.004$&$-6.031\pm0.179$\\
$B_2$& $0.172\pm0.007$&$-5.665\pm0.267$&$B_{s2}$& $0.169\pm0.006$&$-5.765\pm0.252$\\  

\end{tabular}
 \end{ruledtabular}
\end{table}

From the comparison of the slopes in Tables~\ref{tab:rtj},
\ref{tab:rtn} we see that the $\alpha$ values are systematically larger than
the $\beta$ ones. The ratio of their mean values is about 1.4 both for
the charmed and bottom mesons. This value of the ratio is slightly
larger than the one obtained in our recent \cite{lregge} calculations
of the light meson masses, where $\alpha/\beta$ was found to be in
average about 1.3.

We can combine the results of our current calculation 
performed without using the heavy quark $1/m_Q$ expansion with our
previous analysis \cite{hlmass} which was based on such an
expansion in order to analyze the pattern of $P$-levels. As a result
we get the following picture. In the heavy quark limit $m_Q\to\infty$ the
$P$-wave mesons form two heavy quark spin multiplets with light-quark
total angular momentum $j=1/2$ ($0^+, 1^+$) and $j=3/2$
($1^+,2^+$). Masses of the levels with $j=1/2$ are heavier than of the
ones with $j=3/2$. Therefore we have inversion of $P$-levels in the
infinitely heavy quark limit. When we switch on the $1/m_Q$
corrections we get spin splittings in these multiplets and mixing of
the $1^+$ states. Moreover the
levels from these multiplets begin to overlap. This tendency is
further strengthened when the nonperturbative approach in $1/m_Q$
is used. We see from Tables~\ref{tab:csmm},
\ref{tab:bsmm} that there are significant overlaps of the levels
resulting from these multiplets, especially in the charm
sector. However this more sophisticated approach confirms our previous
conclusion that the remnants of the inversion remain in both bottom
and charmed meson spectra. For all considered heavy-light mesons it is
found that the heavier $P_1$ state, which has the main contribution
from the $j=1/2$ multiplet (see above), has the heaviest mass, which is
even higher (by a few MeV for charmed mesons and by almost 30 MeV for
bottom mesons) than the mass of the $^3P_2$ state from the $j=3/2$
multiplet. 

Experimentally complete sets of $1P$-wave meson candidates are known
in the charm sector. In the bottom sector masses of only narrow states
originating from the $j=3/2$ heavy quark spin multiplet are known
reliably. There are some indications of the broad $j=1/2$ states both
of bottom ($0^+$) and bottom-strange ($1^+$) mesons, but additional
confirmation is needed. We find good agreement of our predictions for
$1P$ wave states with available data except for the masses of
 $D^*_{s0}(2317)$ and $D_{s1}(2460)$ mesons. These two charmed-strange
meson states have anomalously low masses which are even lower than the
experimentally observed masses of the corresponding charmed $D^*_0(2400)$
and $D_1(2427)$ mesons. Our model predictions for the masses of the
$1P$-wave $0^+$ and $1^+$ states are almost 200 MeV and 110 MeV higher
than the measured  masses of $D^*_{s0}(2317)$ and $D_{s1}(2460)$
mesons.  Such phenomenon is very hard to understand
within the quark-antiquark picture for these states. Most of the
explanations available in the literature are based on some very
specific fine tuning of the model parameters. The influence of such tuning
on the spectroscopy of other mesons, which are well described in the
framework of the conventional approach, is not well understood. It is
probable that these mesons could have an exotic nature and the genuine
quark-antiquark $P$-wave charmed-strange $0^+$ and $1^+$ states
have higher masses above the $DK$ and $D^*K$ thresholds and are,
therefore, broad. 
We find that  the 
$D^*_{s0}(2317)$ and $D_{s1}(2460)$ mesons do not lie on the corresponding
Regge trajectories. This can be an additional indication of their
anomalous nature. All other experimentally observed $1P$-wave
states match well their trajectories.

Our model suggests that  $D_{s1}(2700)$ and $D^*(2637)$ mesons are the first radial
excitations ($2^3S_1$) of the vector charmed-strange and charmed
mesons. Figures \ref{fig:dstar_j}, \ref{fig:dsstar_j} and
\ref{fig:d_n}, \ref{fig:ds_n} show that they lie on the corresponding
Regge trajectories both in  the $(J,M^2)$ and $(n_r,M^2)$ planes. 

Recent experimental observation \cite{ds3040} that $D_{sJ}^*(2860)$
decays to both $DK$ and $D^*K$ indicates that this state should have
natural parity.  In our model natural parity states $1^-$ ($1^3D_1$)
and $3^-$ ($1^3D_3$) have masses which exceed the experimental value by
about 50 and 100 MeV, respectively. In Ref.~\cite{cfn} it was argued
that from the point of view of decay rates the $3^-$ assignment is
favored. However the measurement of the branching ratios of the
$D_{sJ}^*(2860)$ decay into $D^*K$ to the branching ratio of the decay
into $DK$ differs from the theoretical expectations~\cite{cfn} by three standard
deviations \cite{ds3040}. From Fig.~\ref{fig:dsstar_j} we see that
this state does not fit well to the corresponding Regge trajectory. 

On the other hand, the state $D_{sJ}(3040)$, recently observed by
BaBar \cite{ds3040}  in the $D^*K$ mass spectrum, has a mass coinciding
within errors with the mass of the $1^+$ ($2P_1$) state predicted by our
model (see Table~\ref{tab:csmm}). This state nicely fits to the
daughter Regge trajectory in Fig.~\ref{fig:ds_j}. 

\section{Conclusions}
\label{sec:concl}

The mass spectra of charmed and bottom mesons were calculated in the
framework of the QCD-motivated relativistic quark model. The dynamics
of both light ($q=u,d,s$) and heavy ($Q=c,b$) quarks was treated fully
relativistically without application of either nonrelativistic $v/c$
or heavy quark $1/m_Q$ expansions. The results found in the
nonperturbative in $1/m_Q$ approach confirm the
conclusion, previously obtained withing the heavy quark
expansion up to the first order in Ref.~\cite{hlmass}, that the remnants
of the inversion of the $1P$-levels remain. The final level ordering is rather
complicated, but the higher $1^+$ state is always heavier than the
$2^+$ state. 

We calculated the masses of ground, orbitally and radially excited
heavy-light mesons up to rather high excitations. This allowed us to
construct the Regge trajectories both in $(J,M^2)$ and $(n_r,M^2)$
planes. It was found that they are almost linear, parallel and
equidistant. Most of the available experimental data nicely fit to
them. Exceptions are the anomalously light $D^*_{s0}(2317)$,
$D_{s1}(2460)$ and  $D_{sJ}^*(2860)$ mesons, which masses are 
100-200 MeV lower than various model predictions. The masses of the
charmed-strange $D^*_{s0}(2317)$, $D_{s1}(2460)$ mesons almost coincide
or are even lower than the masses of the partner charmed $D^*_0(2400)$
and $D_1(2427)$ mesons.  These states thus could have an
exotic origin. It will be very important to find the bottom counterparts
of these states in order to reveal their nature.  

\vspace*{1cm}

\acknowledgements
The authors are grateful to  V. Matveev,   M. M\"uller-Preussker,
V. Savrin, D. Shirkov, P. Uwer and M. Wagner for support and discussions.  
This work was supported in part by  the Deutsche
Forschungsgemeinschaft under contract Eb 139/4-1,
the Russian Science Support Foundation 
(V.O.G.) and the Russian Foundation for Basic Research (RFBR), grant
No.08-02-00582 (R.N.F. and V.O.G.).

\end{document}